\documentclass[times,twocolumn]{aastex61}

\usepackage{graphicx}
\usepackage{subfigure}
\usepackage{natbib}
\usepackage{color}
\usepackage{tabularx}
\usepackage{bm}
\usepackage{threeparttable}
\newcommand{\thetae}{\theta_{\rm E}}
\newcommand{\pie}{\pi_{\rm E}}

\newcommand{\kb}{KMT-2016-BLG-1397}
\newcommand{\an}{\theta_{*}}

\DeclareGraphicsExtensions{.pdf,.png,.jpg}

\shorttitle{\kb}
\shortauthors{Zang et al.}

\begin{document}

\title{KMT-2016-BLG-1397\MakeLowercase{b}: KMTNet-only Discovery of a microlens giant planet}

\author[0000-0001-6000-3463]{Weicheng Zang}
\affiliation{Physics Department and Tsinghua Centre for Astrophysics, Tsinghua University, Beijing 100084, China}

\author{Kyu-Ha Hwang}
\affiliation{Korea Astronomy and Space Science Institute, Daejon 34055, Republic of Korea}

\author{Hyoun-Woo Kim}
\affiliation{Korea Astronomy and Space Science Institute, Daejon 34055, Republic of Korea}

\author{Andrew Gould}
\affiliation{Korea Astronomy and Space Science Institute, Daejon 34055, Republic of Korea}
\affiliation{Department of Astronomy, Ohio State University, 140 W. 18th Ave., Columbus, OH 43210, USA}
\affiliation{Max-Planck-Institute for Astronomy, Königstuhl 17, 69117 Heidelberg, Germany}

\author{Tianshu Wang}
\affiliation{Physics Department and Tsinghua Centre for Astrophysics, Tsinghua University, Beijing 100084, China}

\author{Wei Zhu}
\affiliation{Canadian Institute for Theoretical Astrophysics, University of Toronto, 60 St George Street, Toronto, ON M5S 3H8, Canada}

\author{Shude Mao}
\affiliation{Physics Department and Tsinghua Centre for Astrophysics, Tsinghua University, Beijing 100084, China}
\affiliation{National Astronomical Observatories, Chinese Academy of Sciences, A20 Datun Rd., Chaoyang District, Beijing 100012, China}
\affiliation{Jodrell Bank Centre for Astrophysics, Alan Turing Building, University of Manchester, Manchester M13 9PL, UK}

\author{Michael D. Albrow}
\affiliation{University of Canterbury, Department of Physics and Astronomy, Private Bag 4800,
Christchurch 8020, New Zealand}

\author{Sun-Ju Chung}
\affiliation{Korea Astronomy and Space Science Institute, Daejon 34055, Republic of Korea}
\affiliation{Korea University of Science and Technology, 217 Gajeong-ro, Yuseong-gu, Daejeon 34113,
Korea}

\author{Cheongho Han}
\affiliation{Department of Physics, Chungbuk National University, Cheongju 28644, Republic of Korea}

\author{Youn Kil Jung}
\affiliation{Korea Astronomy and Space Science Institute, Daejon 34055, Republic of Korea}

\author{Yoon-Hyun Ryu}
\affiliation{Korea Astronomy and Space Science Institute, Daejon 34055, Republic of Korea}

\author{In-Gu Shin}
\affiliation{Harvard-Smithsonian Center for Astrophysics, 60 Garden St., Cambridge, MA 02138, USA}

\author{Yossi Shvartzvald}
\affiliation{IPAC, Mail Code 100-22, Caltech, 1200 E. California Blvd., Pasadena, CA 91125, USA}

\author{Jennifer C. Yee}
\affiliation{Harvard-Smithsonian Center for Astrophysics, 60 Garden St., Cambridge, MA 02138, USA}

\author{Sang-Mok Cha}
\affiliation{Korea Astronomy and Space Science Institute, Daejon 34055, Republic of Korea}
\affiliation{School of Space Research, Kyung Hee University, Yongin 17104, Republic of Korea}

\author{Dong-Jin Kim}
\affiliation{Korea Astronomy and Space Science Institute, Daejon 34055, Republic of Korea}

\author{Seung-Lee Kim}
\affiliation{Korea Astronomy and Space Science Institute, Daejon 34055, Republic of Korea}
\affiliation{Korea University of Science and Technology, 217 Gajeong-ro, Yuseong-gu, Daejeon 34113,
Korea}

\author{Chung-Uk Lee}
\affiliation{Korea Astronomy and Space Science Institute, Daejon 34055, Republic of Korea}
\affiliation{Korea University of Science and Technology, 217 Gajeong-ro, Yuseong-gu, Daejeon 34113,
Korea}

\author{Dong-Joo Lee}
\affiliation{Korea Astronomy and Space Science Institute, Daejon 34055, Republic of Korea}

\author{Yongseok Lee}
\affiliation{Korea Astronomy and Space Science Institute, Daejon 34055, Republic of Korea}
\affiliation{Korea University of Science and Technology, 217 Gajeong-ro, Yuseong-gu, Daejeon 34113, Korea}

\author{Byeong-Gon Park}
\affiliation{Korea University of Science and Technology, 217 Gajeong-ro, Yuseong-gu, Daejeon 34113,
Korea}

\author{Richard W. Pogge}
\affiliation{Max-Planck-Institute for Astronomy, Königstuhl 17, 69117 Heidelberg, Germany}

\begin{abstract}
We report the discovery of a giant planet in the KMT-2016-BLG-1397 microlensing event, which was found by The Korea Microlensing Telescope Network (KMTNet) alone. The time scale of this event is $t_{\rm E} = 40.0\pm0.5$~days and the mass ratio between the lens star and its companion is $q = 0.016\pm0.002$. The planetary perturbation in the light curve is a smooth bump, resulting in the classical binary-lens/binary-source (2L1S/1L2S) degeneracy. We measure the $V-I$ color of the (putative) two sources in the 1L2S model, and then effectively rule out the binary source solution. The finite-source effect is marginally detected. Combined with the limits on the blend flux and the probability distribution of the source size normalized by the Einstein radius $\rho$, a Bayesian analysis yields the lens mass $M_{\rm L} = 0.45_{-0.28}^{+0.33}~M_\odot$, at distance of $D_{\rm L} = 6.60_{-1.30}^{+1.10}$~kpc. Thus the companion is a super-Jupiter of a mass $m_{p} = 7.0_{-4.3}^{+5.2}~M_{J}$, at a projected separation $r_\perp = 5.1_{-1.7}^{+1.5}$~AU, indicating that the planet is well beyond the snow line of the host star.

\end{abstract}

\section{Introduction}

Since \cite{Shude1991} and \cite{Andy1992} proposed that a search for microlensing of the Galactic bulge stars may lead to a discovery of the extrasolar planetary systems, more than 70 extrasolar planets have been detected by gravitational microlensing\footnote{\url{http://exoplanet.eu/}}. Although relatively few in number, microlensing plays a unique role among planet discovery methods and is complementary to other detection methods \citep{Mao2012, Gaudi2012}. Microlensing probes the planet population beyond the snow line where radial velocity and planetary transit surveys have lower sensitivity. In addition, microlensing planetary systems are distributed at various Galactocentric distances, and therefore a statistical study of them can reveal the effect of different stellar environments (bulge vs. disk) on the planet frequency \citep{Novati2015,Matthewbulge,Zhu2017spitzer}. 

The typical Einstein timescale $t_{\rm E}$ for microlensing events is $\sim20$ days, so a cadence of $\Gamma\sim1~{\rm day}^{-1}$ is sufficient to discover them. However, for planetary signals with characteristic timescales $t_{p} \sim t_{\rm E}\sqrt{q}\to5(q/10^{-4})^{1/2}$ hr \citep{Andy1992} (where $q$ is the planet-host mass ratio), higher cadence is required to characterize the planetary signal. That is $\Gamma \sim 1~{\rm hr}^{-1}$ would be required to discover ``Neptunes'' and $\Gamma \sim 4~{\rm hr}^{-1}$ would be required to detect Earths \citep{Henderson2014}. The Microlensing Observations in Astrophysics (MOA, one 1.8 m telescope equipped with 2.4 ${\rm deg}^2$ camera at New Zealand, \citealt{MOA2016}) and Optical Gravitational Lensing Experiment (OGLE, one 1.3 m telescope equipped with 1.4 ${\rm deg}^2$ camera at Chile, \citealt{OGLEIV}) were the first to conduct wide-area, high-cadence surveys toward the Galactic bulge. Their $\Gamma = 1-4~{\rm hr}^{-1}$ cadences enable the detection of both microlensing events and microlensing planets without the need for follow-up observations \citep[e.g.,][]{OB120406}. 

The Korea Microlensing Telescope Network (KMTNet, \cite{KMT2016}) consists of three 1.6~m telescopes equipped with 4 ${\rm deg}^2$ cameras at CTIO (Chile), SAAO (South Africa) and SSO (Australia). Currently, a total of (3, 7, 11, 3) fields are observed at cadences $\Gamma = (4, 1, 0.4, 0.2)~ {\rm hr}^{-1}$, making it sensitive to planets with masses extending from Jupiter-mass \citep[e.g.,][]{OB150954} to Earth-mass \citep[e.g.,][]{OB161195}. KMTNet has detected planetary perturbations in more than a dozen events since 2015, including four for which the planetary perturbations were only securely detected by KMTNet \citep{KB160212, OB170173, OB170373,OB171552}. However, in most cases, the events themselves were discovered by the OGLE Early Warning System \citep{Udalski2013, Udalski1994} and the MOA \citep{Bond2001} group. 

\cite{Kim2018a, kmtk2, kmt2016data} developed a new microlensing event-finder algorithm for completed events and applied it to the 2016 data, finding 2163 events (1856 ``clear microlensing $+$ 307 ``possible microlensing''), including 861 KMT-only events. Among 70 KMT-only/${\it K2}$C9 events, \cite{KB160212} announced the discovery of a substellar companion by high-cadence data ($\Gamma = 4~{\rm hr}^{-1}$) in the KMT-2016-BLG-0212 microlensing event, with two possible solutions (low-mass brown-dwarf or sub-Neptune companion). This event has a baseline of $I_{\rm base} \sim 19.2$ and then rises to a peak $I_{\rm peak} \sim 18.8$, with a 4-hour anomaly $I_{\rm anom} \sim 18.2$, making it difficult to be detected by real-time alert systems.

Here we report the analysis of the KMT-only planetary event \kb\, with planet/host mass ratio $q = 0.016 \pm 0.002$. The smooth bump around ${\rm HJD}^{\prime}$ = 7513 (${\rm HJD}^{\prime}$ = HJD - 2450000) results in the classical degeneracy between binary-lens/single-source (2L1S) and single-lens/binary-source (1L2S) solutions \citep{Gaudi1998}. We resolve the 2L1S/1L2S degeneracy by measuring the $V-I$ color of the (putative) two sources, and thus effectively rule out the 1L2S solution. The paper is structured as follows. In Section \ref{obser}, we describe the observations of KMTNet. We then fit the data with a binary-lens model in Section \ref{BL} and check the binary-source solution in Section \ref{BS}. In Section \ref{lens} we estimate the physical parameters of the planetary system. Finally, our conclusions and the implications of our work are given in Section \ref{dis}.



\label{intro}

\section{Observations and event recognition}\label{obser}


\kb\ was located at equatorial coordinates $(\alpha, \delta)_{\rm J2000}$ = (18:10:39.51, $-24$:51:27.86), corresponding to galactic coordinates $(\ell,b)=(6.32,-2.80)$. It therefore lies in the KMTNet BLG31 field, monitored with a cadence of $\Gamma = 0.4~{\rm hr}^{-1}$. As mentioned in Section \ref{intro}, KMTNet observations are carried out with three 1.6~m telescopes at CTIO (Chile, KMTC), SAAO (South Africa, KMTS) and SSO (Australia, KMTA). The great majority of data were taken in the standard $I$ band, with occasional observations made in the standard $V$ band. 


\kb\ was originally recognized as ``clear microlensing" by the event-finding algorithm \citep{Kim2018a,kmt2016data}, because its light curve has obvious microlens features and the algorithm found that the $\Delta\chi^2$ improvement relative to a flat line is 15759, which is a robust evidence compared to the $\Delta\chi^2 = 1000$ threshold. The source star was identified as BLG31K0508.007617 in the KMT Dophot catalog.  


The anomaly was discernible in the pySIS $I$-band light curve (extracted using pySIS software package, \citealt{pysis}) by human review. This review comprised an inspection of automated pySIS light curves in the neighborhood of the event candidate and 2016-2017 joint difference imaging analysis (DIA) light curves (extracted by a customized pipeline based on \citealt{Alard1998} and \citealt{Wozniak2000}). In general, the quality of the pySIS light curves are better than those from the DIA pipeline, mainly because pySIS finds the true position of the microlens source, but also because it employs a more accurate point-spread function. The final pySIS light curves that we used in the analysis were reduced by hand for optimal photometry. \textbf{We also calibrated the pySIS light curves to standard Cousins $I$ magnitude by matching the baseline flux to the nearest star in OGLE-III catalog \citep{Udalski2013}.} We note that there were $\sim50$ obvious outliers at baseline ($>5\sigma$ deviations from all the single-lens, binary -lens and binary source models), which we eliminated before modeling the data. In addition, we find that the camera of KMTC was stuck in the BLG31 field for some unknown reasons on the night of ${\rm HJD}^{\prime}$ = 7545.XX, resulting in six points being taken within 13 minutes, four with somewhat reduced flux. These 6 points only have influence in the model with parallax parameters, leading to a weak but unuseful constraint ($\pie<2$). Therefore, we also eliminated all six KMTC data points on that night.



\section{Binary-lens modeling}\label{BL}
\subsection{Grid search and global minima}
We first search the space of `Standard' binary-lens solutions. The model has seven geometric parameters to calculate the magnification, $A(t)$. These include three point-lens point-source (PLPS) parameters $t_0$, $t_{\rm E}$ and $u_0$: the time of the maximum magnification, the Einstein radius crossing time and the impact parameter in units of the angular Einstein radius $\thetae$, respectively \citep{Paczynski1986}. We also need: the source size normalized by the Einstein radius, $\rho$; the binary mass ratio, $q$; the projected separation between the binary components normalized to the Einstein radius, $s$; and the angle between the source trajectory and the binary axis in the lens plane, $\alpha$. The event is observed as a change in flux $F(t)$ at the location of the event
\begin{equation}
    f(t) = f_{\rm s} A(t) + f_{\rm b},
\end{equation}
where $f_{\rm s}$ is the flux of the source star being lensed, and $f_{\rm b}$ represents any blended flux that is not lensed\footnote{\textbf{We choose 18 as the magnitude zeropoint. Thus, the magnitude of the source can be derived by $I_{\rm s} = 18 - 2.5 * \log_{10}(f_{\rm s}$)}}. The two linear parameters, $F_{\rm s}$ and $F_{\rm b}$ will be different for each observatory and each filter. We use the advanced contour integration code, \texttt{VBBinaryLensing}\footnote{http://www.fisica.unisa.it/GravitationAstrophysics/VBBinaryLensing.htm}, to compute the magnification $A(t)$ (See \cite{Bozza2010} for more details). 

We undertake a grid search on parameters ($\log s, \log q, \alpha$), with 20 values equally spaced between $-1\leq\log s \leq1$, $0^{\circ}\leq \alpha \leq 360^{\circ}$, and 40 values equally spaced between $-4\leq \log q \leq0$. For each set of ($\log s, \log q, \alpha$), we fix $\rho=0.001$ and find the minimum $\chi^2$ by a downhill\footnote{We use a function based on the Nelder-Mead simplex algorithm from the SciPy package. See \url{https://docs.scipy.org/doc/scipy/reference/generated/scipy.optimize.fmin.html\#scipy.optimize.fmin}} approach on the remaining three parameters ($t_0, u_0, t_E$). As a result, we find that there are three distinct minima, with ($\log s, \log q, \alpha) \sim$ ($-0.4, -0.3, 72.0$) or ($0.2, -2.2, 324.0$) or ($-0.1, -2.2, 324.0$). We note that the last two solutions are the $s \leftrightarrow s^{-1}$ degenerate solutions. 

\subsection{Best-fit model}
If finite-source effects are measured in the light curve, we should include the limb-darkening effect. The form of the limb-darkening law we use is 
\begin{equation}
    S_\lambda(\mu) = \bar{S}_\lambda \left[1 - \Gamma_{\lambda}(1 - \frac{3}{2}\mu)]\right],
\end{equation}
where $\bar{S}_\lambda$ is the mean surface brightness of the source, $\mu$ is the cosine of the angle between the normal to the stellar surface and the line of sight, and $\Gamma_{\lambda}$ is the limb-darkening coefficients at wavelength $\lambda$. From the color analysis in Section \ref{cmd}, we infer $\Gamma_{I} = 0.3696$ for $I$ band, and $\Gamma_{V} = 0.5265$ for $V$ band.  



\begin{figure}[htb] 
    \includegraphics[width=\columnwidth]{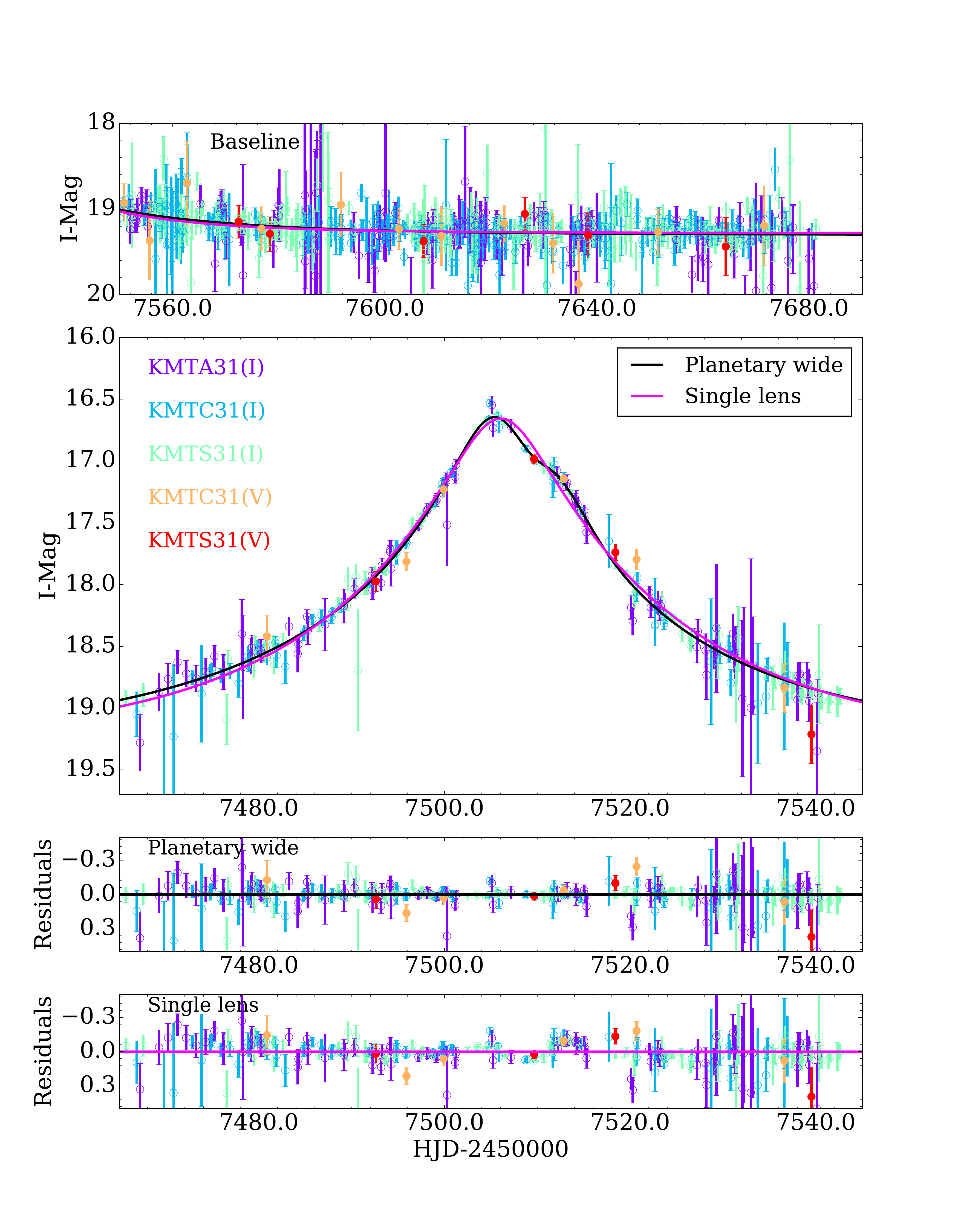}
    \caption{The light curve of event \kb. The top panel shows the baseline of the data. In the second panel, the black and magenta lines are the light curves for the best-fit planetary wide and single-lens model, respectively. The last two panels show the residuals from the  best planetary wide and single-lens model, respectively. The red and yellow dots are $V$ band data points, the circles are $I$ band data points. The light curve and data has been calibrated to standard $I$-band magnitude.}
    \label{fig:lc}
\end{figure}

\begin{figure}[htb] 
    \includegraphics[width=\columnwidth]{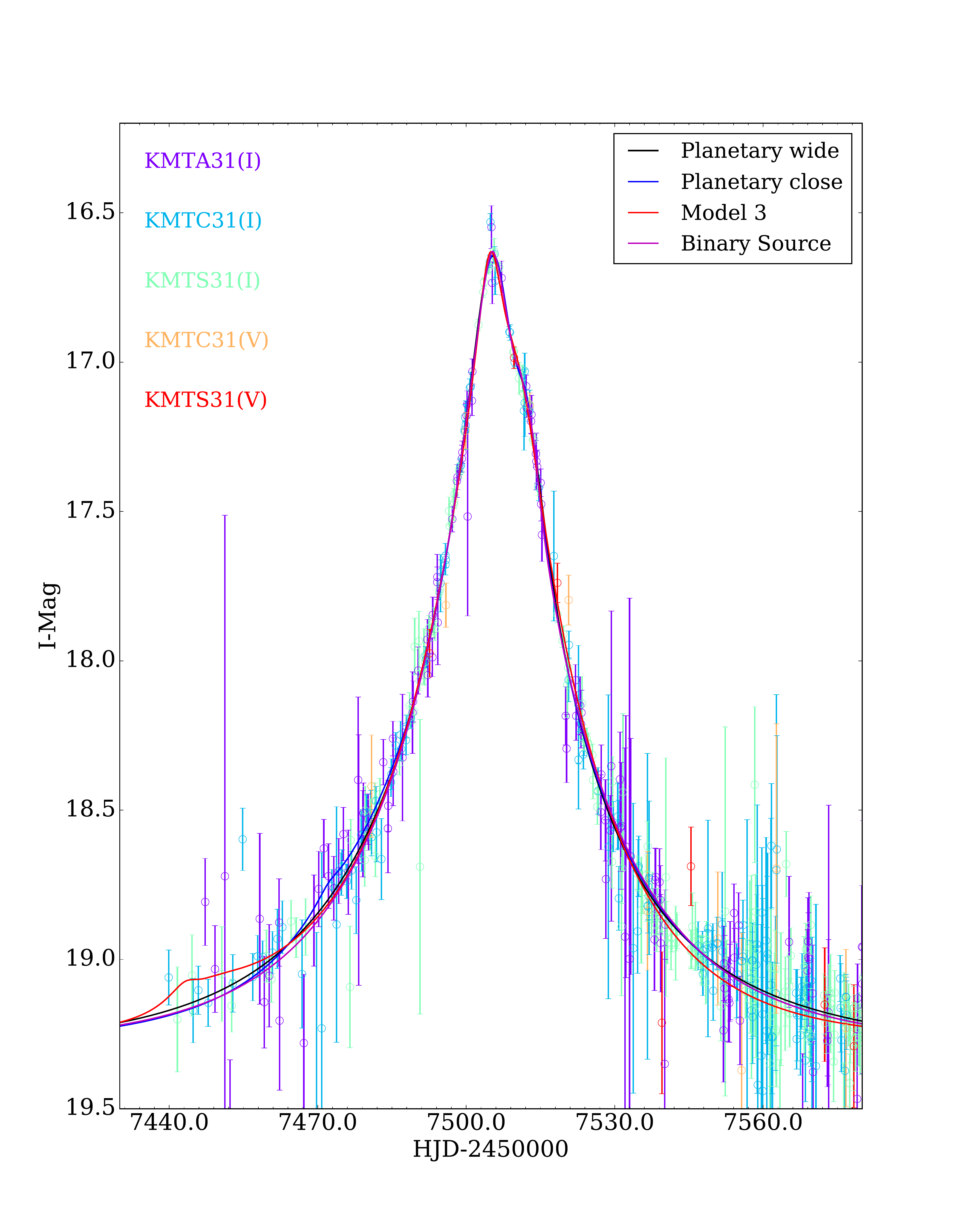}
    \caption{The light curves of the three binary-lens models and the binary source model.}
    \label{fig:com}
\end{figure}

\begin{figure}[htb] 
    \includegraphics[width=\columnwidth]{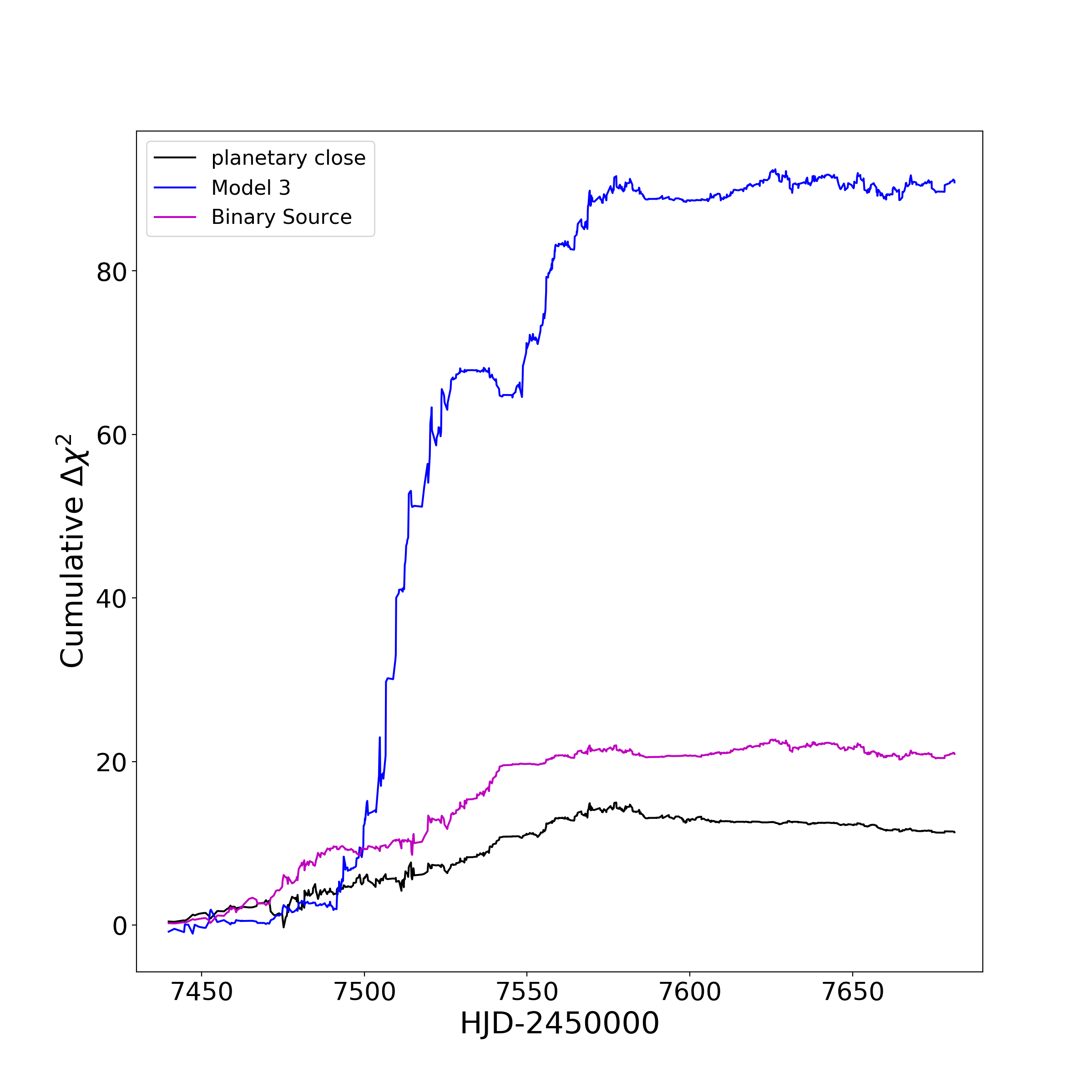}
    \caption{Cumulative distribution of $\chi^2$ differences ($\Delta\chi^2 = \chi^2_{\rm model} - \chi^2_{\rm wide}$) between the three models (planetary close, binary lens model 3 and binary source) and the planetary wide model. It is clear that the $\chi^2$ differences are not mainly from outliers.}
    \label{fig:res}
\end{figure}

\begin{figure*}[htbp]
    \centering
    \subfigure{
    \begin{minipage}{10cm}
    \centering
    \includegraphics[width=\columnwidth]{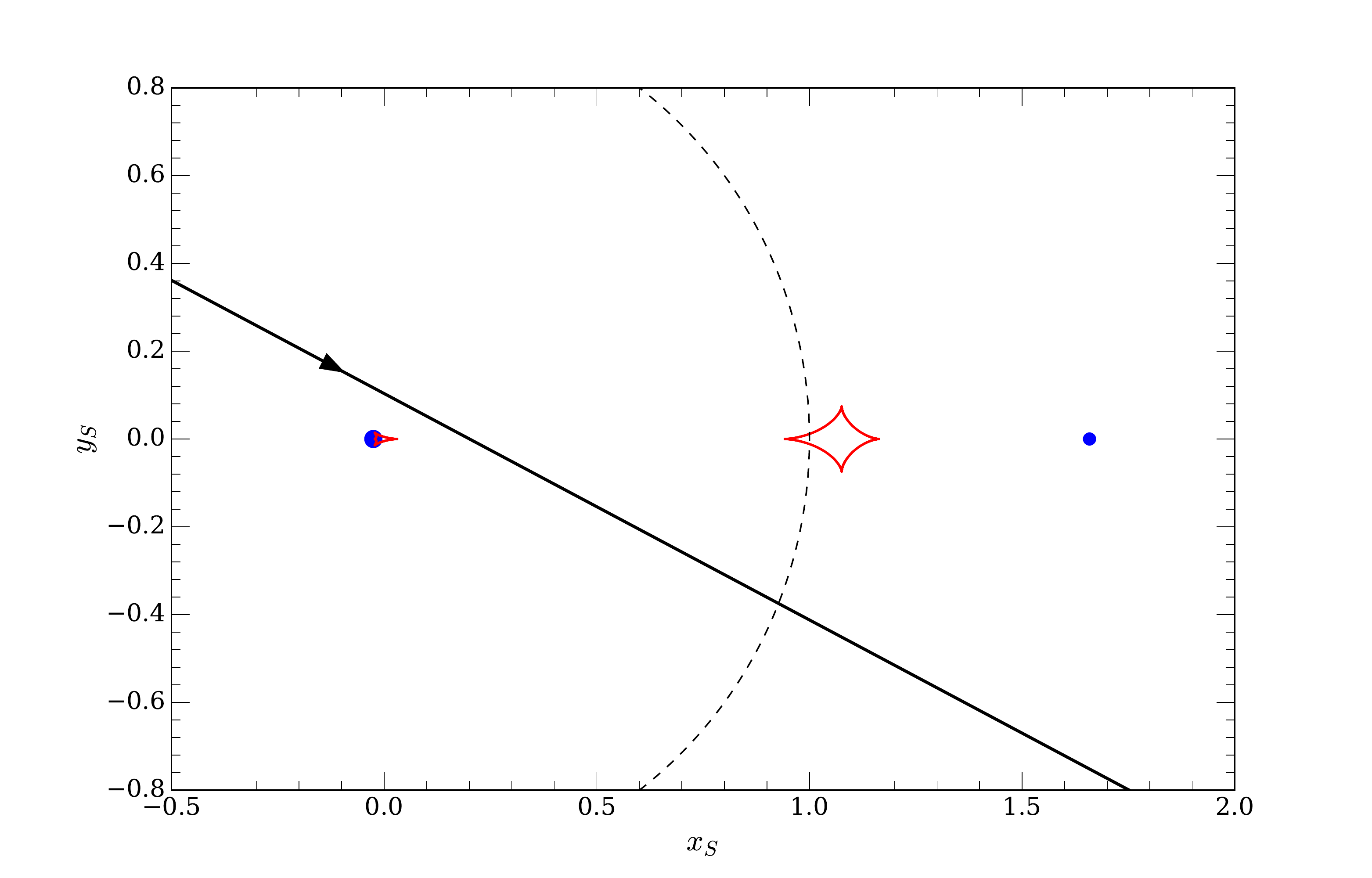}
    \end{minipage}
    }
    
    \subfigure{
    \begin{minipage}{10cm}
    \centering
    \includegraphics[width=\columnwidth]{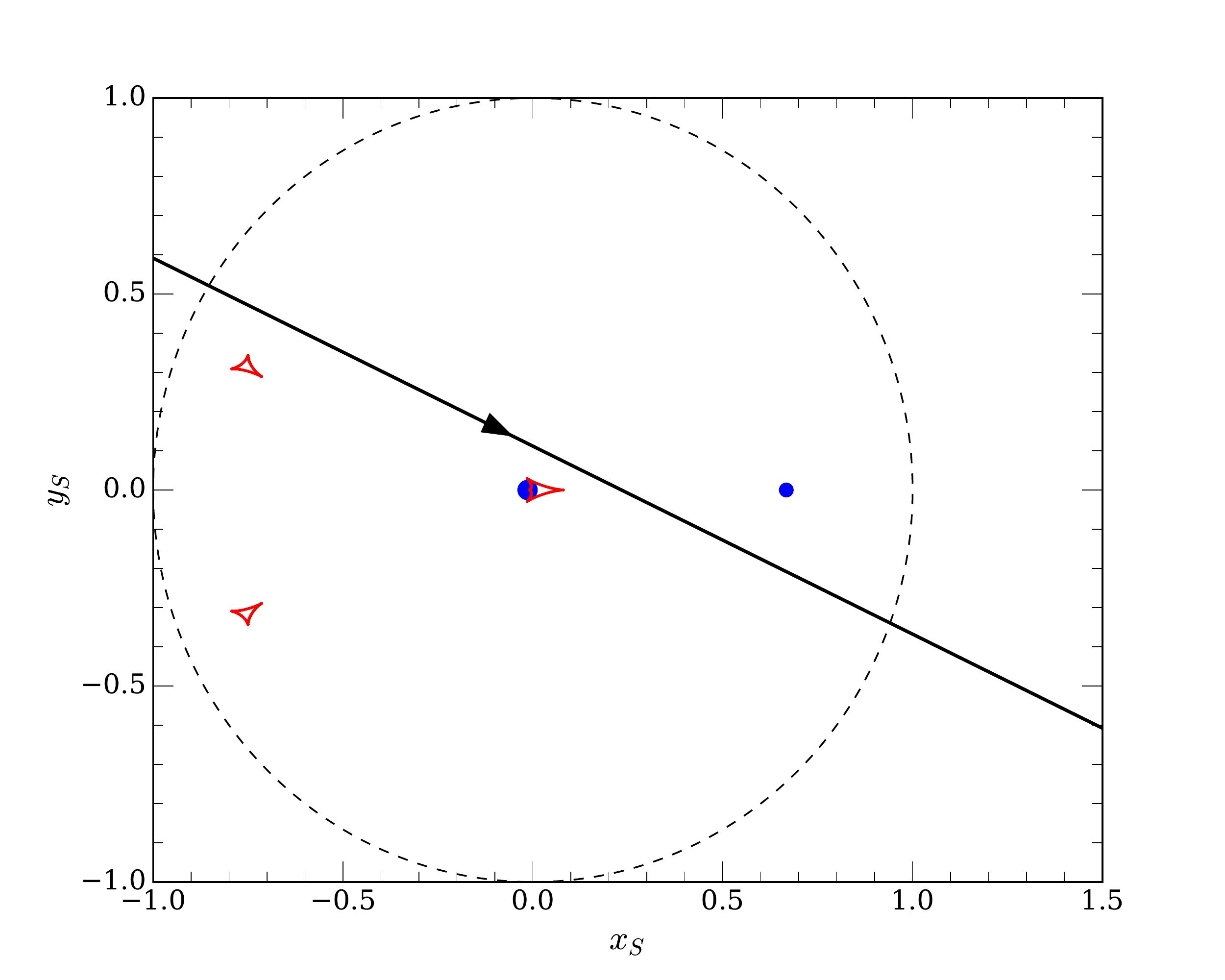}
    \end{minipage}
    }
    \subfigure{
    \begin{minipage}{10cm}
    \centering
    \includegraphics[width=\columnwidth]{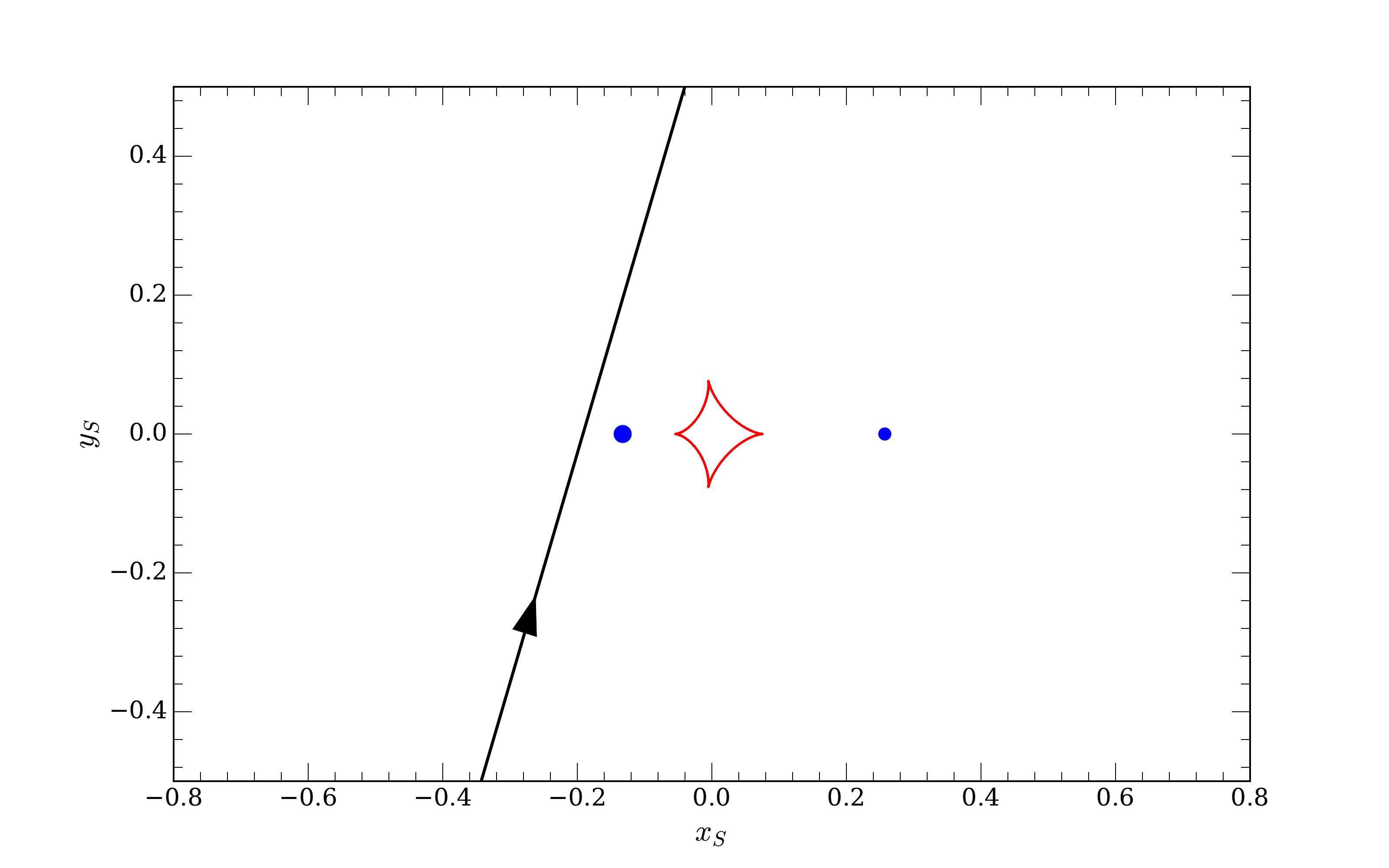}
    \end{minipage}
    }
    \caption{Geometries of the best-fit planetary wide ($s>1$, upper panel), planetary close ($s<1$, middle panel) and binary lens model 3 ($s=0.39, q=0.51$, lower panel). In each panel, the red closed curves are the caustics. The two blue dots are the positions of the two components. The solid line is the trajectory of the source, and the arrow indicates the direction of source motion. The axes are in units of the Einstein angle $\thetae$, and the dashed line is the angular Einstein ring of the lens system.}
    \label{fig:cau}
\end{figure*}

Setting the initial parameters as those of the two minima, we then employ Markov Chain Monte Carlo (MCMC) $\chi^2$ minimization using the \texttt{emcee} ensemble sampler \citep{emcee} with free ``standard model" parameters to search for the best-fit model. Finally, the solution $(s, q) = (1.68\pm0.05, 0.016\pm0.002)$ is the best-fit model, while its $s \leftrightarrow s^{-1}$ degenerate solution $(s, q) = (0.66\pm0.03, 0.025\pm0.004)$ is disfavored by $\Delta\chi^2$ = 11. Another solution $(s, q) = (0.39\pm0.01, 0.51\pm0.01)$ is disfavored by $\Delta\chi^2 > 80$ as well as its large negative blended flux $f_b = -0.407\pm0.029$. Thus we only adopt the planetary wide solution ($s > 1$) in the following analysis. In addition, we also fit the data with a PLPS model. The $\chi^2$ improvement between the planetary wide and the PLPS model is 349. The best-fit parameters of the binary-lens and the PLPS model are shown in Table \ref{table:parm1}, the best-fit model curves for planetary wide and PLPS models are shown in Figure \ref{fig:lc}, and the best-fit model curves and their cumulative distribution of $\chi^2$ difference for the three binary-lens solutions are shown in Figure \ref{fig:com} and \ref{fig:res} respectively. We also show the geometries of the three binary-lens solutions in Figure \ref{fig:cau}. 


Our modeling gives an upper limit on the source size normalized by the Einstein radius, $\rho < 0.046$ ($3\sigma$ level). The best-fit model has $\rho = 0.029$, but the data are also consistent with a point-source model at $\sim1.6\sigma$ level ($\Delta \chi^2 = 2.6$). We also try including the microlens parallax effect in modeling, but it does not improve the fit significantly ($\Delta \chi^2 = 4$). The upper limit of the microlens parallax as the $3\sigma$ level is about 2.1, which gives no useful constraints.  



\section{Binary-source modeling}\label{BS}
\cite{Gaudi1998} first pointed out the potential degeneracy between 2L1S and 1L2S solutions. The first practical example of this was the low-mass planetary event OGLE-2005-BLG-390, for which the 1L2S model was qualitatively consistent with the data, but was quantitatively rejected at $\Delta\chi^2>50$ \citep{OB05390}.  Nevertheless, there have been several subsequent events with plausible planetary solutions that proved to be binary-source (1L2S) or even triple-source (1L3S) events \citep{MB12486, OB151459, OB160733}. Therefore, we consider the possibility that the small bump around ${\rm HJD}^{\prime}$ = 7513 is caused by a single-lens event with a binary source.


In the case of binary-source events, the total magnification is the superposition of two 1L1S events generated by the individual source stars \citep{Shude1991,Griest1992, Han_BS}. The total magnification $A_\lambda$ at wavelength $\lambda$ is:
\begin{equation}
   A_\lambda = \frac{A_1F_{1,\lambda}+A_2F_{2,\lambda}}{F_{1,\lambda}+F_{2,\lambda}} = \frac{A_{1}+q_{F,\lambda}A_{2}}{1+q_{F,\lambda}},
\end{equation}

\begin{equation}
    q_{F,\lambda} = \frac{F_{2,\lambda}}{F_{1,\lambda}},
\end{equation}
where $A_i$ ($i = 1, 2$) is the magnification of each source with flux $F_i$.



We fit pySIS $I$-band data with a binary-source model using MCMC, and the best-fit parameters are shown in Table \ref{table:parm1}. The binary-source model is disfavored by  $\chi^2_{\rm 1L2S} - \chi^2_{\rm 2L1S}$ = 20.5. Although this result strongly disfavors the binary-source model, we nevertheless seek additional confirmation by evaluating the color evolution of the light curve.

\cite{Shude1991} first proposed that a difference in colors of two sources will make a color change during binary source events. \cite{Gaudi1998} noted that a binary-source event and a planetary event can be distinguished by the color difference expected for the two sources of different luminosities. For example, \cite{MB11322} used this method to confirm the planetary interpretation of MOA-2011-BLG-322, for which the putative two sources are G- and K-type main sequence stars, while \cite{OB151459} demonstrated the correctness of the 1L3S solution for OGLE-2015-BLG-1459 due to strong color evolution. To obtain the color of the two sources, we perform a special set of pyDIA reductions of the data (i.e., different from the pySIS reductions from the main light-curve analysis) because the pyDIA light-curve photometry is tied to the same system as field-star photometry (instrumental magnitude scale). We then find that the unlensed instrumental $I$ magnitude of the two sources would be $16.04\pm0.07$ and $18.98\pm0.22$ (See Table \ref{table:parm2} for the lensing parameters). Applying the parameters of the red clump in Section \ref{cmd}, the brighter source is about 2.1 magnitude below the red clump. Thus it is a subgiant and could be consistent with a broad range of colors. In the 1L2S model, it is about $-0.35$ mag bluer than the red clump. So the brighter source would be a bluish subgiant. Regarding the fainter source, it is about $5.1$ magnitude below the red clump and hence would be an early K dwarf. Thus, it should be about $-0.15$ mag bluer than the red clump \citep{starparm}, which is inconsistent with the color that we obtain in the 1L2S model, of $-0.71\pm0.19$ mag. Hence, the color analysis confirms the rejection of the 1L2S solution based on $\Delta\chi^2$.




\section{Physical parameters}\label{lens}

\subsection{Color-Magnitude Diagram}\label{cmd}
To further constrain the lens properties, we estimate the angular radius $\theta_*$ of the source by placing the source on a color-magnitude diagram (CMD) \citep{Yoo2004}. We construct the CMD by stars within a $120^{\prime\prime}$ square centered on the source position using KMTC data. We estimate the red clump to be $(V-I, I)_{\rm cl} = (2.41\pm0.02, 13.91\pm0.06)$ and find that the source is $\Delta(V - I) = -0.44\pm0.03$ bluer and $\Delta I = 2.23\pm0.08$ fainter than the red clump. From \cite{Bensby2013,Nataf2013}, we find that the intrinsic color and de-reddened brightness of the red clump are $(V-I, I)_{\rm cl,0} = (1.06, 14.27)$. Thus, the intrinsic color and de-reddened magnitude of the source are $(V-I, I)_{\rm S,0} = (0.62\pm0.03, 16.50\pm0.08)$. We then employ the color/surface-brightness relation in \cite{Adams2018}, and finally find 
\begin{equation}\label{angular}
    \an =  1.50 \pm 0.06~\mu {\rm as}.
\end{equation}

For the binary-lens models, we obtain a $3\sigma$ upper limit $\rho<0.046$. This allows us to set a lower limit on the angular Einstein radius $\theta_{\rm E}$,
\begin{equation}
    \thetae > \frac{\theta_*}{\rho_{\rm upper}} = 0.033~\text{mas}.
\end{equation}

We can also estimate the effective temperature $T_{\rm eff} = 6000~$K of the source by using the color-temperature relation in \cite{Houdashelt2000}. Using ATLAS models and assuming a metallicity of [M/H] = 0.0, a microturbulence parameter of 1 km/s and a surface gravity of $\log g$ = 4.0, we obtain the linear limb-darkening coefficients $u_I = 0.4679$ for $I$ band,  $u_V = 0.6252$ for $V$ band \citep{Claret2011}. This, when combined with the transformation formula in \cite{Fields2013}, yields the corresponding limb-darkening coefficients $\Gamma_{I} = 0.3696$, $\Gamma_{V} = 0.5625$. 

\subsection{Bayesian analysis}\label{Bayesian}
For a lensing object, the total mass $M_{\rm L}$ and distance $D_{L}$ is related to observables by
\begin{equation}
    M_{\rm L} = \frac{\thetae}{{\kappa}\pie};~
    D_{\rm L} = \frac{\mathrm{au}}{\pie\thetae + \pi_{\rm S}},
    \label{eq:mass}
\end{equation}
where $\kappa \equiv 4G/(c^2\mathrm{au}) = 8.144$ mas$/M_{\odot}$ is a constant, $\thetae$ is the angular Einstein radius, $\pie$ is the microlensing parallax~\citep[e.g.,][]{Gould2000} and $\pi_{\rm S} = \mathrm{au}/D_{\rm S}$ is the source parallax \citep{Gouldpies1992,Gouldpies2004}. For KMT-2016-BLG-1397, neither $\thetae$ nor $\pie$ is unambiguously measured. However, combining the timescale $t_{\rm E}$, the limits of blend flux and the probability distribution of $\rho$, we can use Bayesian analysis to estimate the mass and distance of the lens system.

\begin{figure*}[htbp]
    \centering
    \subfigure{
    \begin{minipage}{8cm}
    \centering
    \includegraphics[width=\columnwidth]{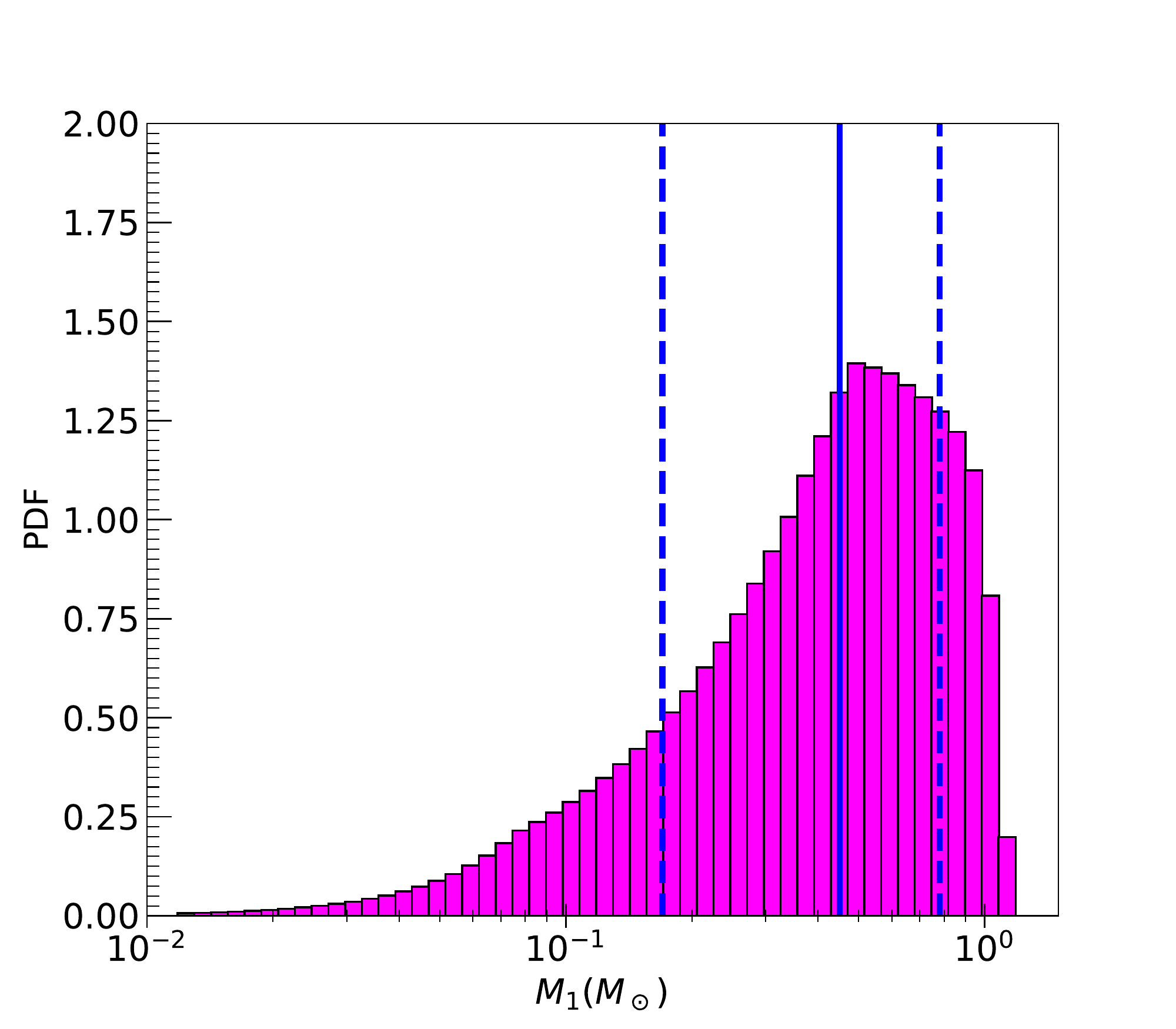}
    \end{minipage}
    }
    \subfigure{
    \begin{minipage}{8cm}
    \centering
    \includegraphics[width=\columnwidth]{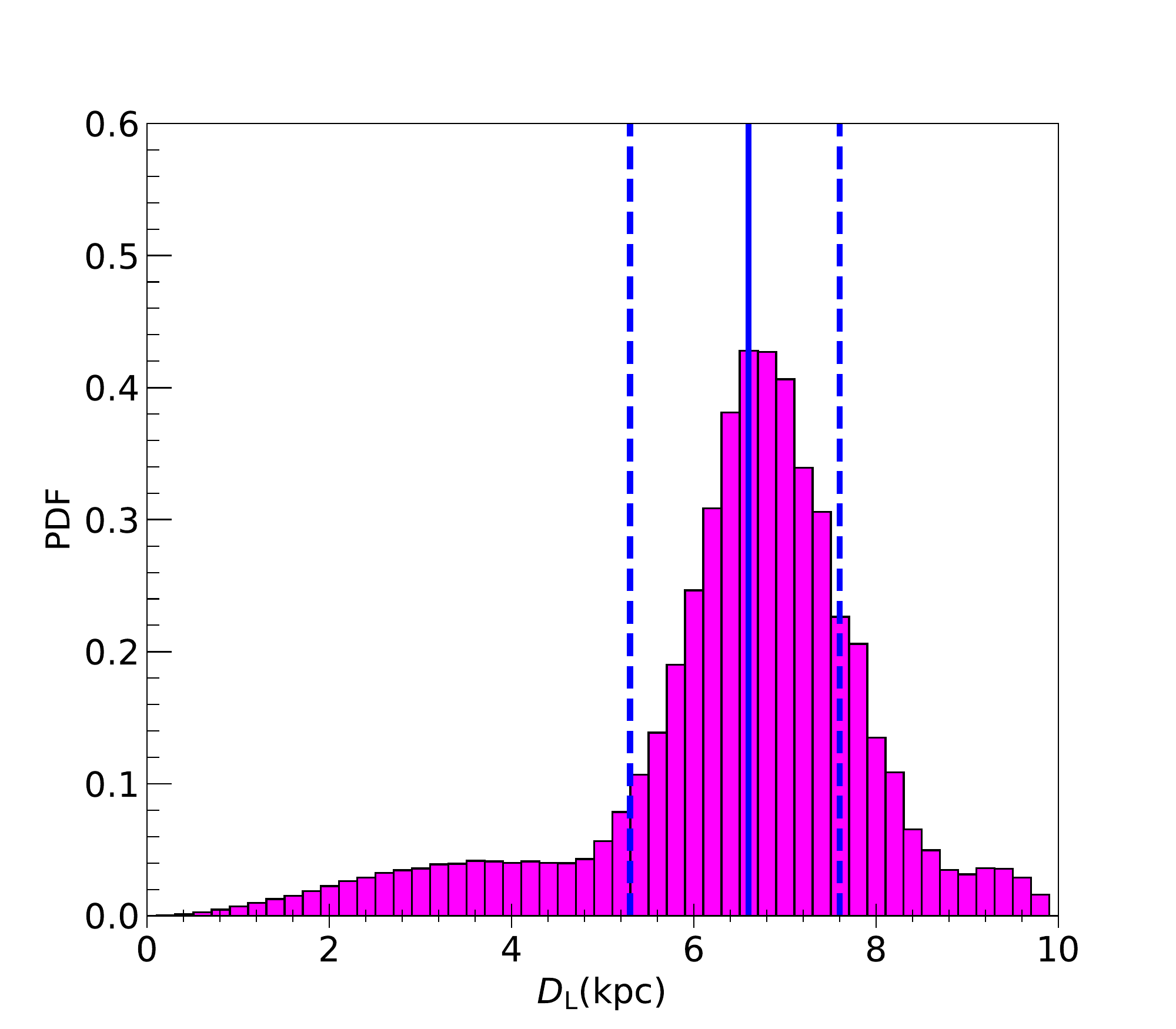}
    \end{minipage}
    }
     \subfigure{
    \begin{minipage}{8cm}
    \centering
    \includegraphics[width=\columnwidth]{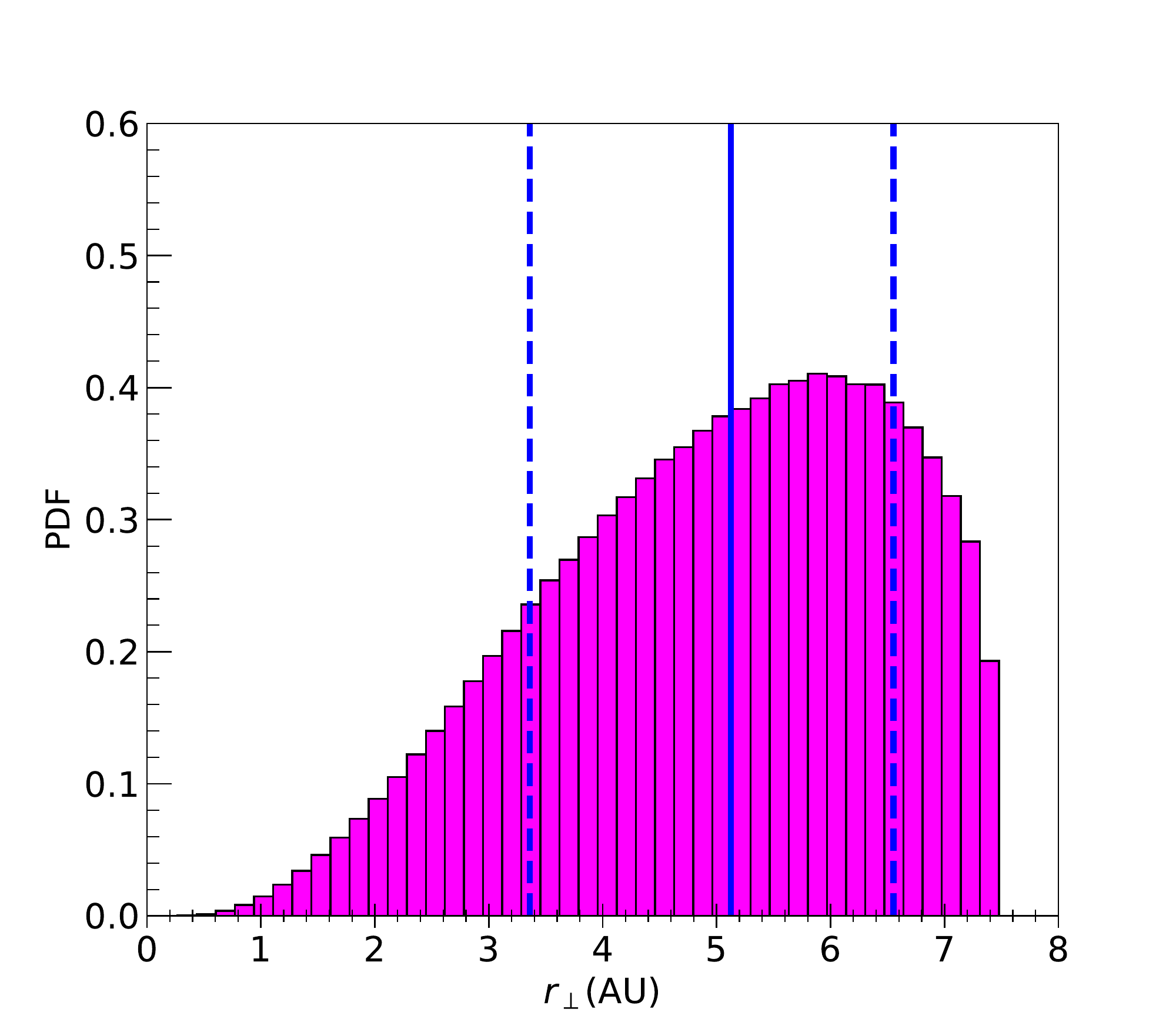}
    \end{minipage}
    }

    \caption{Bayesian posterior probability density distributions (PDFs) for the lens mass $M_{\rm L}$ (the upper left panel), the lens distance $D_{\rm L}$ (the upper right panel), and the projected separation $r_\perp$ of the planet (the lower panel). In each panel, the blue solid vertical line represents the median value and the two blue dashed lines represent 16th and 84th percentiles of the distribution.}
    \label{fig:baye}
\end{figure*}

We applied the Galactic model and Bayesian model described in \cite{Zhu2017spitzer} to randomly draw lensing events. We weight all events by exp($-\Delta\chi^2(\rho)/2$), where $\Delta\chi^2(\rho)$ represents the $\chi^2$ difference relative to the minimum $\chi^2$ for the lower envelope of the ($\chi^2$ vs.~$\rho$) diagram derived from the MCMC. In addition, the blend flux is $I_{\rm b,inst} = 17.92 \pm 0.15$, To be conservative, we set an upper limit of the blend flux to be $I_{\rm b,limit} = 17.47~(3\sigma)$. We adopt the mass-luminosity relation \citep{OB171130},
\begin{equation}
    M_I = 4.4 - 8.5 \log{\frac{M}{M_{\odot}}},
\end{equation}
where $M_I$ is the absolute magnitude in $I$-band. We reject trial events for which the lens distance obeys 
\begin{equation}
    M_I + 5 \log{\frac{D_{\rm L}}{10 {\rm pc}}} < I_{\rm b,limit} - \Delta I_{\rm cl},
\end{equation}
where $\Delta I_{\rm cl} = I_{\rm cl} - I_{\rm cl,0} = -0.36$. The resulting posterior distributions of the lens mass $M_{\rm L}$, distance $D_{\rm L}$, lens-source relative proper motion $\mu$ and the projected separation $r_\perp$ of the planet are shown in Figure \ref{fig:baye}. We find that the lens mass is $M_{\rm L} = 0.45_{-0.28}^{+0.33}~M_\odot$.  The uncertainties are the $68\%$ probability range about the median of the probability distribution, which we take as the most likely value. Thus the planet mass is $M_{\rm planet} = 7.0_{-4.3}^{+5.2}~M_{J}$. The lens distance is $D_{\rm L} = 6.60_{-1.30}^{+1.10}$~kpc, and the planet is at a projected separation $r_\perp = 5.1_{-1.7}^{+1.5}$~AU from the host star, indicating that the planet is well beyond the snow line of the host star (assuming a relation $r_{\rm SL} = 2.7(M/M_{\odot})$~{\rm AU}, \citealt{snowline}).

\section{Discussion and Conclusion}\label{dis}
We have reported the discovery and analysis of an extrasolar planet KMT-2016-BLG-1397b that was detected by gravitational microlensing. The event was found by the KMTNet survey alone. This is one of the first KMT-only planets and has a planetary mass ratio $q = 0.016\pm0.002$. The other candidate KMT-only planet, KMT-2016-BLG-0212, however, has two classes of solutions, characterized by low-mass brown-dwarf ($q = 0.037$) and sub-Neptune ($q < 10^{-4} $) companions, respectively. In contrast to most of the microlensing planets detected by high-cadence surveys ($\Gamma \geq 1~{\rm hr}^{-1}$), the planetary perturbation of \kb\ was detected by the KMTNet survey with a $\Gamma = 0.4~{\rm hr}^{-1}$ low cadence. This is the third planet observed by KMTNet in such low-cadence areas (the other two events are OGLE-2016-BLG-0263 \citealt{OB160263} and OGLE-2016-BLG-1067 \citealt{OB161067}). However, in the other two cases, the planetary perturbations were covered by MOA's $\Gamma > 1~{\rm hr}^{-1}$ data. Thus \kb\ showcases the power of the KMTNet with wide sky coverage and observations from three sites. This event also shows that multiple-color observations are quite important because the additional color information allowed us to decisively resolve the 2L1S/1L2S degeneracy.

\begin{figure}[htb] 
    \includegraphics[width=\columnwidth]{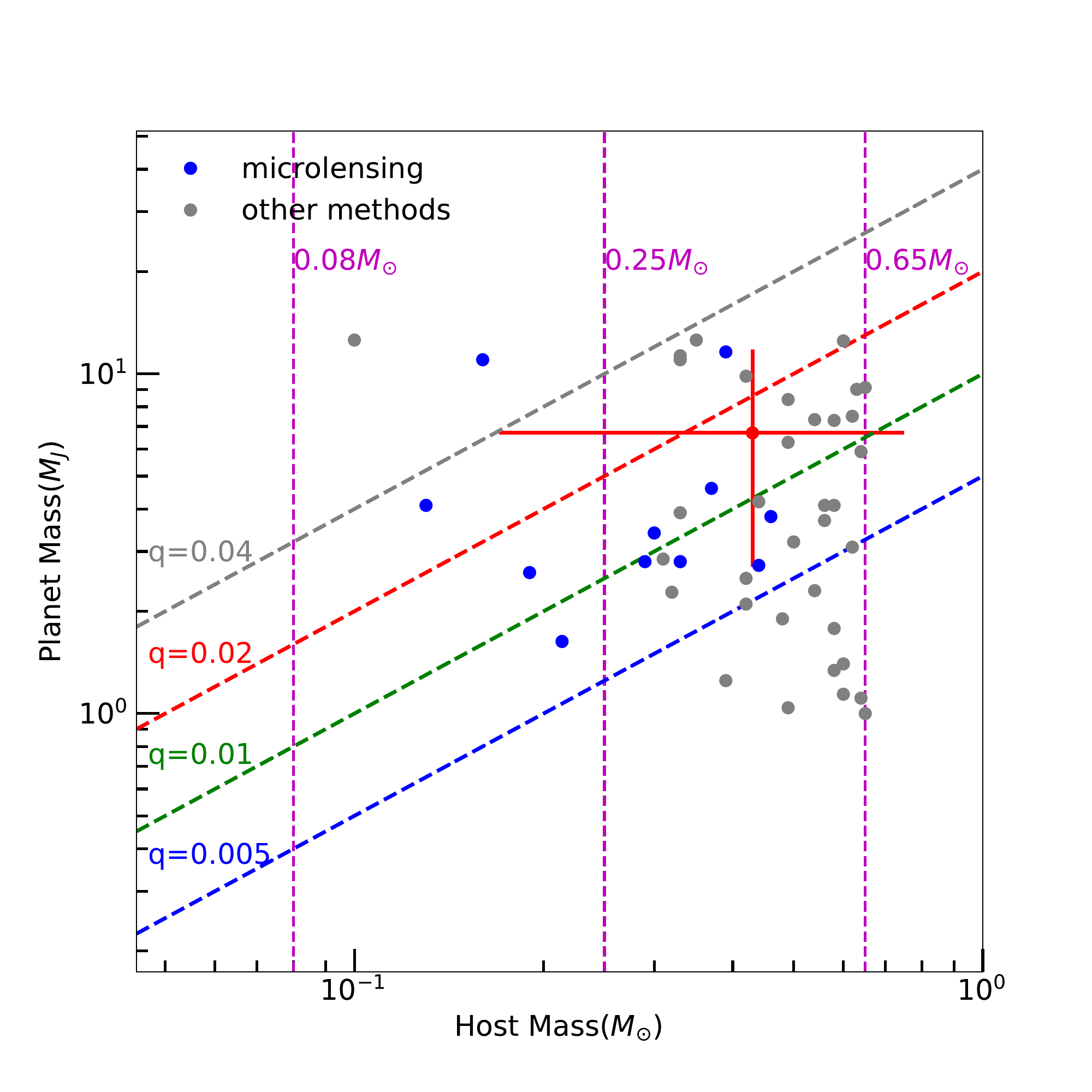}
    \caption{Mass distribution of known M dwarf/super Jupiter system. We select $0.08M_{\odot} < M_{\rm host} < 0.65M_{\odot}, 1.0M_{J} < M_{p} < 13.5M_{J}$ from \url{http://exoplanetarchive.ipac.caltech.edu}. The red dot is \kb. The planets discovered by microlensing are marked by blue dots, while those found by other methods are marked by grey dots. The magenta vertical dashed lines represent the conventional star/brown-dwarf boundary ($0.08 M_{\odot}$), the conventional M dawrf/K dwarf boundary ($0.65 M_{\odot}$) and the rough low-mass star boundary ($0.25 M_{\odot}$).}
    \label{fig:MM}
\end{figure}

The finite-source effect is only marginally detected while the microlens parallax gives no useful constraints. Our Bayesian analysis yields the lens mass $M_{\rm L} = 0.45_{-0.28}^{+0.33}~M_\odot$, and the planet mass $m_{p} = 7.0_{-4.3}^{+5.2}~M_{J}$, which consists of an M dwarf orbited by a super-Jupiter mass planet ($\sim 70\%$ probability). In the core-accretion theory, massive planets around M dwarfs should be rare because of insufficient material for forming planets \citep{Laughlin2004, Ida2005}. \kb\ is the twelfth case for which a massive planet orbits an M dwarf detected by microlensing\footnote{OGLE-2005-BLG-071 \citep{OB050071, OB050071D}, MOA-2009-BLG-387 \citep{MB09387}, MOA-2010-BLG-073 \citep{MB10073}, MOA-2011-BLG-322 \citep{MB11322}, OGLE-2012-BLG-0406 \citep{OB120406,OB120406Y} OGLE-2008-BLG-355 \citep{OB080355}, OGLE-2015-BLG-0954 \citep{OB150954,OB150954moa}, OGLE-2013-BLG-1761 \citep{OB131761}, MOA-2016-BLG-227 \citep{MB16227}, OGLE-2016-BLG-0263 \citep{OB160263}, OGLE-2017-BLG-1140 \citep{OB171140}}, which demonstrates that such giant planets around low-mass stars are quite common. In Figure \ref{fig:MM}, we compare the mass distribution of this lens system to previously known M dwarf/super Jupiter systems ($0.08M_{\odot} < M_{\rm host} < 0.65M_{\odot}, 1.0M_{J} < M_{p} < 13.5M_{J}$). The figure shows that gravitational microlensing is a powerful method of detecting M dwarf/super Jupiter systems, and it finds 12/47 of the known such systems. In addition, most of (4/5) systems that have a very low mass host star ($0.08M_{\odot} < M_{\rm host} < 0.25M_{\odot}$) are detected by microlensing. This can be easily understood because very low mass stars are extremely faint and microlensing is the only method that does not rely on the light from the star-planetary system. Thus the distinctive sensitivity regimes of microlensing can improve our understanding of M dwarf/super Jupiter system formation mechanisms.   



\begin{figure*}[htbp]
    \centering
    \subfigure{
    \begin{minipage}{8cm}
    \centering
    \includegraphics[width=\columnwidth]{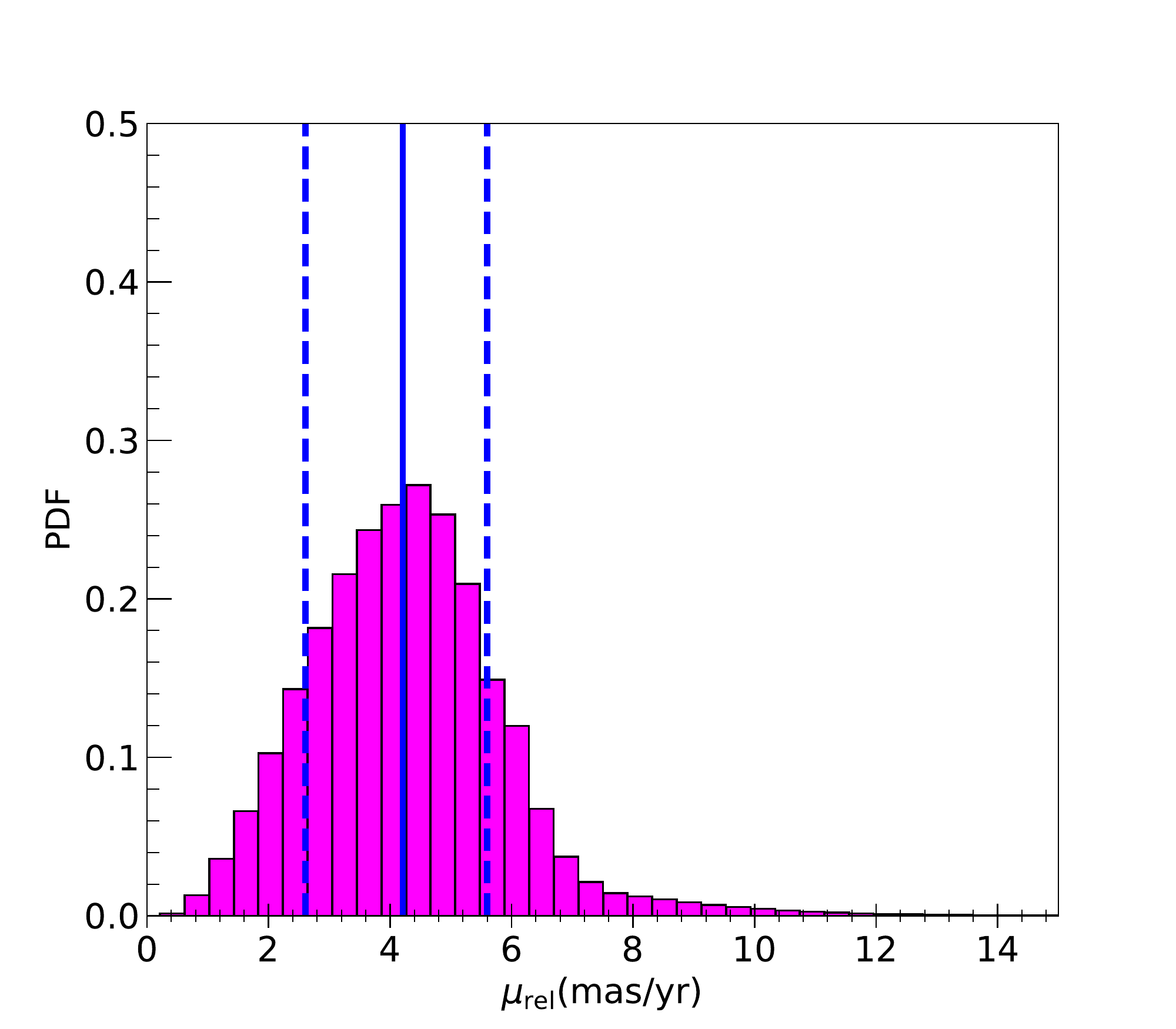}
    \end{minipage}
    }
    \subfigure{
    \begin{minipage}{8cm}
    \centering
    \includegraphics[width=\columnwidth]{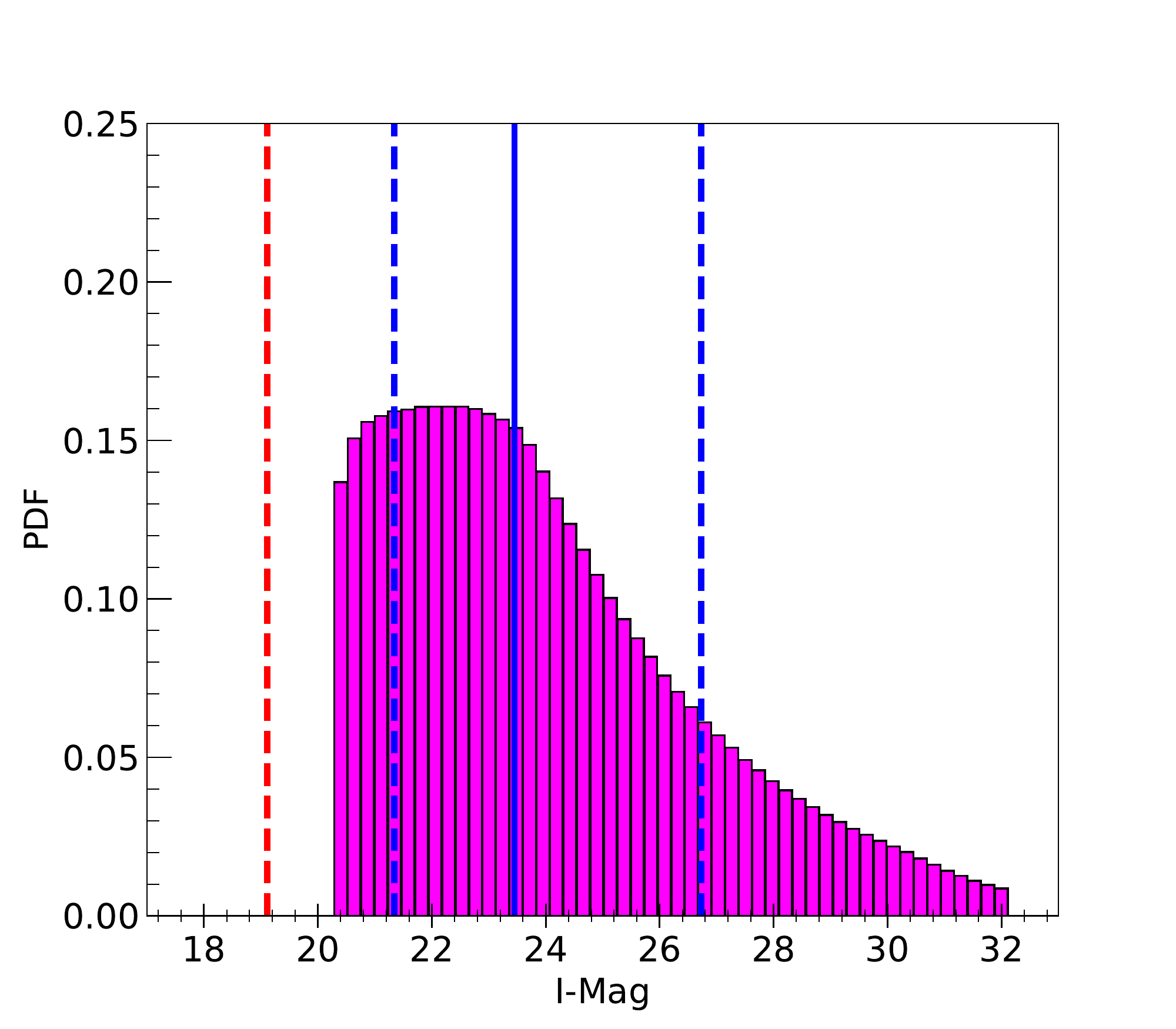}
    \end{minipage}
    }
     \subfigure{
    \begin{minipage}{8cm}
    \centering
    \includegraphics[width=\columnwidth]{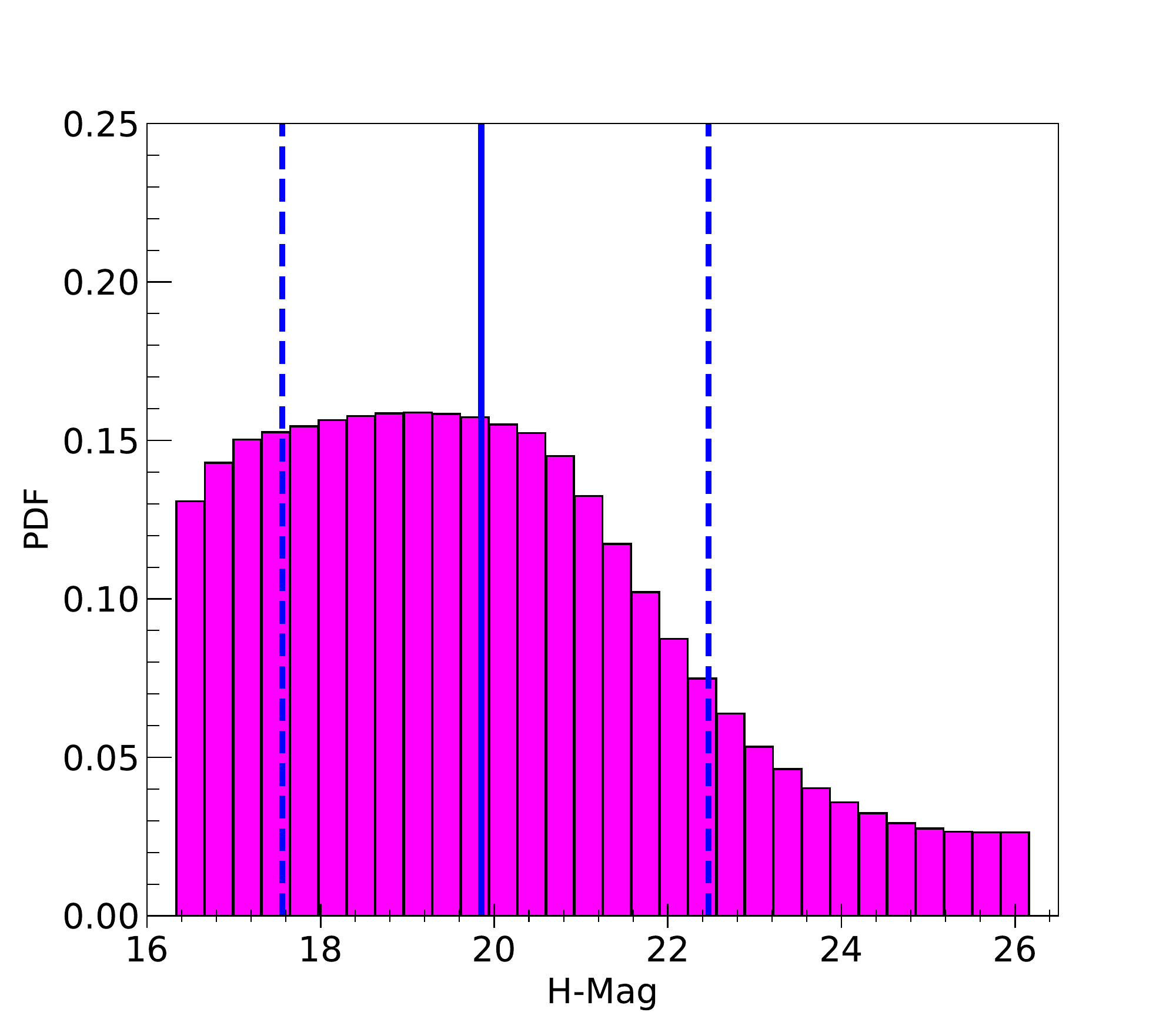}
    \end{minipage}
    }
    \caption{Bayesian posterior probability density distributions (PDFs) for the lens-source relative proper motion $\mu_{\rm rel}$ (the upper left panel), the lens brightness in $I$ band (the upper right panel) and $H$ band (the lower panel). In each panel, the blue solid vertical line represents the median value and the two blue dashed lines represent 16th and 84th percentiles of the distribution. \textbf{In the upper right panel, the red dashed line is the brightness of the source in $I-$band.}}
    \label{fig:pm}
\end{figure*}

\cite{Batista2015} fully resolved the source and lens of OGLE-2005-BLG-169 using Keck adaptive optics (AO) when they were separated by $\sim$ 60~mas, while \cite{Dave2015} resolved them by the Hubble Space Telescope (HST) when they were separated by $\sim$ 48~mas. \textbf{Recently, \cite{Bhattacharya2018} resolved the source and lens of OGLE-2012-BLG-0950 using Keck and the HST when they were separated by $\sim$ 34~mas}. In these case, the source and lens have approximately equal brightness. Our Bayesian analysis provides us with estimates of the lens-source relative proper motion $\mu_{\rm rel} = 4.2_{-1.6}^{+1.4} {\rm mas\,yr^{-1}}$  and the brightness of the lens $I_{\rm L} = 23.45_{-2.11}^{+3.28} , H_{\rm L} = 19.85_{-2.29}^{+2.62}$ (We adopt $A_H = 0.46$ from \citealt{Gonzalez2012}. See Figure \ref{fig:pm}). Because the source is 50-100 times brighter than the lens, it will probably require about 80~mas separation to resolve the source and lens with current instruments. This would require a $\sim20$ year wait. However, the upcoming next generation ($D \sim 30$m class, such as E-ELT, TMT and GMT) telescopes have a resolution $\theta \sim 14(D/30{\rm m})^{-1}$ mas in $H$ band, and the lens-source relative proper motions of the 12 microlensing M dwarf/super Jupiter systems are $\mu > 2$ mas/yr. Thus \kb\ and previously discovered M dwarf/super Jupiter systems can be confirmed/contradicted at first light of AO images on 30m telescopes .


\vspace{6pt}

We thank Tianjun Gan and Chelsea Huang for discussions. This work was partly supported by the National Science Foundation of China (Grant No. 11333003, 11390372 and 11761131004 to SM). This research has made use of the KMTNet system operated by the Korea Astronomy and Space Science Institute (KASI) and the data were obtained at three host sites of CTIO in Chile, SAAO in South Africa, and SSO in Australia. Work by WZ, YKJ, and AG were supported by AST-1516842 from the US NSF. WZ, IGS, and AG were supported by JPL grant 1500811. AG is supported from KASI grant 2016-1-832-01. AG received support from the European Research Council under the European Union’s Seventh Framework Programme (FP 7) ERC Grant Agreement n. [321035].

\bibliography{Zang}

\begin{table*}[htb]
    \centering
    \caption{Lensing Parameters from pySIS data}
    \begin{tabular}{c|c|c|c|c|c}
    \hline
    \hline
    Parameters & Planetary wide(s$>$1) & Planetary close(s$<$1) & Binary lens model 3 & Binary-source & PLPS \\
    \hline
    $t_{0,1}$ - 2450000 (d) & 7506.5(1) & 7506.3(1)  & 7505.65(4) & 7505.2(1) & 7506.04(3) \\
    $t_{0,2}$ - 2450000 (d) & ... & ... & ... & 7512.5(1) & ...\\
    $u_{0,1}$  & 0.090(3) & 0.101(4)  & 0.186(8) & 0.133(7) & 0.18(1)\\
    $u_{0,2}$  & ... & ... & ... & 0.05(2) & ... \\
    $t_E (d)$ & 41.0(5) & 37.1(4)  & 25.5(8) & 35.0(4) & 28(1) \\
    s  & 1.68(6) & 0.66(3) & 0.39(1)  & ... & ... \\
    q  & 0.016(2) & 0.025(4)  & 0.51(1) &  ... & ...\\
    $\alpha$ (rad) & 5.81(1) & 5.84(1) & 1.28(1) & ... & ... \\
    $\rho_{1}$ & $<$0.046 & $<$0.044  & $<$0.048 & 0.11(3) & ... \\
    $\rho_{2}$ & ... & ... & ... & 0.08(3) & ...\\
    $q_{F,I}$ & ... & ... & ...& 0.055(18) & ...\\
    $f_{\rm s}$ & 0.361(7) & 0.384(8) & 0.715(29) & 0.444(8) & 0.661(33)  \\
    $f_{\rm b}$ & $-0.060(7)$ & $-0.081(8)$ & $-0.407(29)$ & $-0.140(8)$ & $-0.360(33)$  \\
    \hline
    $\chi^2/dof$  & 889.8/888 &  900.8/888 & 979.5/888 & 910.3/887 & 1238.9/892 \\
    \hline
    \end{tabular}\\
    \label{table:parm1}
\end{table*}

\begin{table}[htb]
    \centering
    \caption{Binary-Source Parameters from pyDIA data}
    \begin{tabular}{c|c}
    \hline
    \hline
    Parameters & \\
    \hline
    $t_{0,1}$ - 2450000 (d) & 7505.2(1)\\
    $t_{0,2}$ - 2450000 (d) & 7512.6(1) \\
    $u_{0,1}$  & 0.12(1)\\
    $u_{0,2}$  & 0.05(1) \\
    $t_E (d)$ & 35.5(9)\\
    $q_{F,I}$ & 0.063(20)\\
    $q_{F,V}$ & 0.086(39)\\
    $\rho_{1}$ & 0.09(4)\\
    $\rho_{2}$ & 0.07(3)\\
    $I_{s,1}$ & 16.04(7)\\
    $(V-I)_1$ & 2.06(4)\\
    $I_{s,2}$ & 18.98(22)\\
    $(V-I)_2$ & 1.67(19)\\
    \hline
    \end{tabular}\\
    \label{table:parm2}
\end{table}

\begin{figure}[htb] 
    \includegraphics[width=\columnwidth]{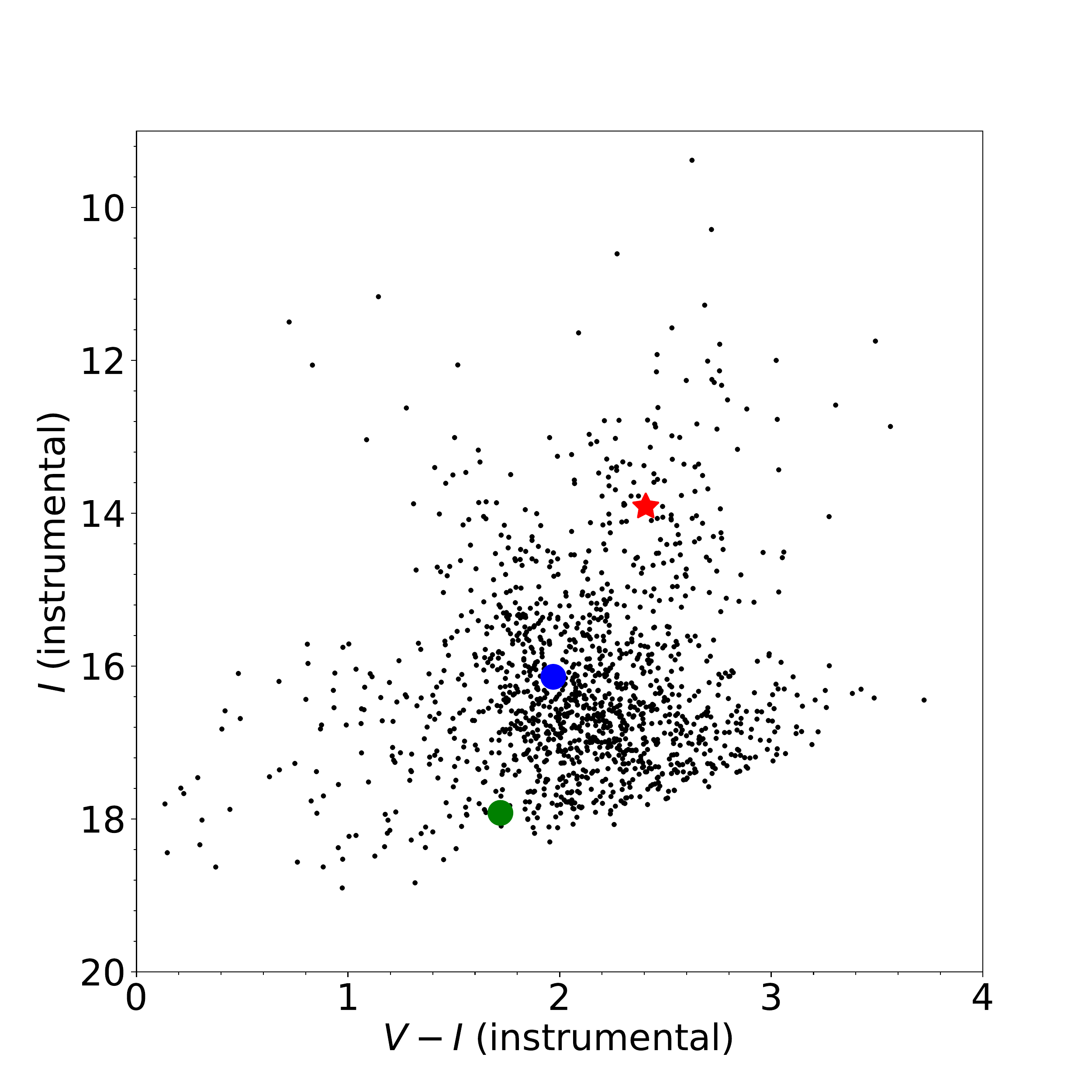}
    \caption{Instrumental color-magnitude diagram of a $120^{\prime\prime}$ square centered on \kb\ (using KMTC data). The red asterisk shows the centroid of the red clump. The blue dot indicates the position of the source, and the green dot is the position of the blended light.}
    \label{fig:cmd}
\end{figure}

\end{document}